\documentclass{appolb}
\usepackage{epsfig}

\newcommand{\lsim}
{\mathrel{\raisebox{-.3em}{$\stackrel{\displaystyle <}{\sim}$}}}

\def\asymp#1%
{\mathrel{\raisebox{-.4em}{$\widetilde{\scriptstyle #1}$}}}

\def\Nequal#1%
{\mathrel{\raisebox{-.5em}{$\stackrel{=}{\scriptstyle\rm#1}$}}}
\newcommand{\dsl}[1]{\not \hspace{-0.7mm}#1}
\def\dsl{\mathpalette\make@slash}
\def\make@slash#1#2{\setbox\z@\hbox{$#1#2$}%
  \hbox to 0pt{\hss$#1/$\hss\kern-\wd0}\box0}

\def\beq{\begin{equation}}
\def\eeq{\end{equation}}
\def\beqar{\begin{eqnarray}}
\def\eeqar{\end{eqnarray}}
\def\barr#1{\begin{array}{#1}}
\def\earr{\end{array}}
\def\bfi{\begin{figure}}
\def\efi{\end{figure}}
\def\btab{\begin{table}}
\def\etab{\end{table}}
\def\bce{\begin{center}}
\def\ece{\end{center}}
\def\nn{\nonumber}
\def\disp{\displaystyle}
\def\text{\textstyle}

\def\ga{\gamma}
\def\de{\delta}

\def\eps{\epsilon}
\def\veps{\varepsilon}

\def\citere#1{\mbox{Ref.~\cite{#1}}}
\def\citeres#1{\mbox{Refs.~\cite{#1}}}

\newcommand{\TeV}{\unskip\,\mathrm{TeV}}
\newcommand{\GeV}{\unskip\,\mathrm{GeV}}
\newcommand{\MeV}{\unskip\,\mathrm{MeV}}

\newcommand{\ri}{{\mathrm{i}}}
\newcommand{\rd}{{\mathrm{d}}}

\newcommand{\Oa}{\mathswitch{{\cal{O}}(\alpha)}}

\newcommand{\M}{{\cal{M}}}

\def\mathswitchr#1{\relax\ifmmode{\mathrm{#1}}\else$\mathrm{#1}$\fi}
\newcommand{\Pf}{\mathswitch  f}

\newcommand{\PW}{\mathswitchr W}
\newcommand{\Pw}{\mathswitchr w}
\newcommand{\PZ}{\mathswitchr Z}

\newcommand{\PH}{\mathswitchr H}
\newcommand{\Pe}{\mathswitchr e}

\newcommand{\Pt}{\mathswitchr t}

\newcommand{\Pep}{\mathswitchr {e^+}}
\newcommand{\Pem}{\mathswitchr {e^-}}

\def\mathswitch#1{\relax\ifmmode#1\else$#1$\fi}

\newcommand{\MW}{\mathswitch {M_\PW}}

\newcommand{\MZ}{\mathswitch {M_\PZ}}
\newcommand{\MH}{\mathswitch {M_\PH}}
\newcommand{\Me}{\mathswitch {m_\Pe}}

\newcommand{\Mt}{\mathswitch {m_\Pt}}

\newcommand{\sw}{\mathswitch {s_\Pw}}
\newcommand{\cw}{\mathswitch {c_\Pw}}

\newcommand{\GF}{\mathswitch {G_\mu}}

\def\solid{\raise.9mm\hbox{\protect\rule{1.1cm}{.2mm}}}
\def\dash{\raise.9mm\hbox{\protect\rule{2mm}{.2mm}}\hspace*{1mm}}

\newcommand{\eetth}{\Pep\Pem\to\Pt\bar\Pt\PH}

\begin{document}
\title{{Electroweak Physics}
\thanks{Presented by A. Denner at the final meeting of the European Network
        ``Physics at Colliders'', Montpellier, France, September 26-27, 2004.}}
\author{W. Hollik
\address{Max-Planck-Institut f\"ur Physik, D-80805 Munich}
\and
A. Accomando, F. del Aguila, M. Awramik,
A. Ballestrero,
J.~van der Bij, W. Beenakker, R. Bonciani, M. Czakon, 
G.~Degrassi,
A. Denner, K. Diener, S.~Dittmaier, A. Ferroglia,
J. Fleischer, A. Freitas, N. Glover, J. Gluza, T. Hahn,
S.~Heinemeyer, S.~Jadach, F.~Jegerlehner,
W.~Kilian, M.~Kr\"amer, J.~K\"uhn, 
E.~Maina, S.~Moretti,
T.~Ohl, C.G.~Papadopoulos, G.~Passarino, R.~Pittau, 
S.~Pozzorini, M.~Roth, T.~Riemann, J.B.~Tausk,
S.~Uccirati, \\
A.~Werthenbach, G.~Weiglein 
\address{NSCR Democritos Athens, CERN,
Univ.~Granada, Univ.~Durham, DESY, Univ.~Edinburgh,
Univ.~Freiburg, Univ.~Karlsruhe, Univ.~Katowice,   
HNINP~Krakow,
Max Planck Institut f\"ur Physik M\"unchen,
Univ.~Nijmegen, Univ.~Roma~3, Univ.~Southampton,
Univ.~Torino,
PSI~Villigen, Univ.~W\"urzburg}
}
\maketitle
\begin{abstract}
Work on electroweak precision calculations and event generators for 
electroweak physics studies at 
current and future 
colliders is summarized.
\end{abstract}

\section{Introduction}

Apart from the still missing Higgs boson, 
the Standard Model (SM) has been impressively confirmed
by successful collider experiments at the particle accelerators
LEP, SLC, and  Tevatron during the last decade with high precision,
at the level of radiative corrections.
Future colliders like the upcoming LHC or an 
$\Pep\Pem$ International Linear Collider (ILC), 
offer an even greater physics potential, 
and in turn represent a 
great challenge for theory
to provide even more precise calculations.
Accurate predictions are necessary not only to increase the level of
precision of SM tests, but also to study the indirect effects of
possible new particles.

With increasing energies many new processes with heavy particle production
will be accessible. Such 
heavy particles are unstable, so that their production leads
to many-particle final states, with e.g.\ four or six fermions.
Predictions for such reactions should be based
on full transition matrix elements, improved by radiative corrections
as far as possible,
and call for proper event generators.

Joint work done within the network is reviewed here; more details can be 
found also in previous studies for future colliders
\cite{LHCbook,Aguilar-Saavedra:2001rg,Baur:2001yp}.

\section{Precision observables and multi-loop calculations}

\subsection{Muon lifetime and $W$--$Z$ mass correlation }

The precise measurement of the muon lifetime, or equivalently of the
Fermi constant $\GF$, sets an important constraint on the SM parameters,
\beq
\GF = \frac{\pi\alpha(0)}{\sqrt{2}\MW^2\sw^2}\,(1+\Delta r),
\eeq
with $\sw^2=1-\cw^2=1-\MW^2/\MZ^2$, where
the quantity $\Delta r$ comprises the radiative corrections to 
muon decay (apart from the photonic corrections in the Fermi model).
Solving this relation
for the W-boson mass $\MW$ yields a precise prediction for $\MW$
that can be compared with the directly measured value.
Recently the full electroweak two-loop calculation has been completed,
with the contributions from closed fermion 
loops~\cite{Freitas:2000gg,Awramik:2003ee}, from
bosonic loops involving virtual Higgs-bosons~\cite{Freitas:2000gg},
and the complete bosonic corrections~\cite{Awramik:2002wn,Awramik:2003ee}.
The two-loop fermionic contributions influence the $\MW$
prediction at the level of $\sim 50\MeV$,
the two-loop bosonic corrections by $1{-}2\MeV$.

The predictions at the two-loop level have been 
further improved by universal
higher-order contributions to the $\rho$-parameter. The 
terms of the order  
${\cal O}(\GF^2\Mt^4\alpha_{\mathrm{s}})$ and
${\cal O}(\GF^3\Mt^6)$ have been obtained  
in \cite{Faisst:2003px}
and were found to change $\MW$ at the level
of $5\MeV$ and $0.5\MeV$, respectively.
The leading three-loop bosonic contribution to the $\rho$-parameter
in the large Higgs mass limit~\cite{Boughezal:2004},
yielding the power $M_{\rm H}^4/M_{\rm W}^4$,
is opposite in sign to the leading two-loop correction
$\sim M_{\rm H}^2/M_{\rm W}^2$. The two terms cancel each other 
for a Higgs boson mass around 480 GeV. This interesting new result 
stabilizes the perturbative expansion and 
makes a strongly interacting Higgs sector very unlikely in view of the
electroweak precision data.  

The prediction for $\MW$,
including the above-mentioned two-loop and leading three-loop effects,
carries, besides the parametric uncertainty,
an intrinsic theoretical uncertainty, which is estimated to be of
$3{-}4\MeV$ \cite{HollikZeuthenprocs,Gw2004}, 
which has to be
compared with the aimed precision of
$7\MeV$ in the $\MW$ determination at a future ILC~\cite{Baur:2001yp}.

\subsection{Precision observables at the Z~resonance}

In order to describe the Z-boson resonance at LEP1 within satisfactory
precision it was possible to parametrize the cross section 
near the resonance in such a way \cite{zresonance,Bardin:1997xq}
that a Born-like form with generalized
``effective'' couplings is convoluted with QED structure functions
modeling initial-state radiation (ISR).
From these effective Z-boson--fermion couplings so-called 
``pseudo-observables'' were derived, such as various asymmetries, the
hadronic Z-peak cross section, partial Z-decay widths, etc.
The precisely calculated  
pseudo-observables are implemented in the programs
{\sc Zfitter} and {\sc Topaz0}~\cite{Bardin:1998nm}.
A critical overview on high-precision physics at the Z~pole, 
in particular focusing on the theoretical uncertainties, can be 
found in \cite{zeuthenMHworkshop}.
The status of precision pseudo-observables 
in the MSSM is summarized 
in~\cite{Heinemeyer:2004gw}. 

Following the formal tree-level like parametrization of the couplings, 
an ``effective weak mixing angle'', 
$\sin^2\theta_{f}^{\rm eff}$, was derived for each fermion species.
Among these parameters the leptonic variable
$\sin^2\theta_{\ell}^{\rm eff}$ plays a particularly important role,
since it is measured with the high accuracy of $1.7\times 10^{-4}$
and is very sensitive to the Higgs-boson mass.
Quite recently,
a substantial subclass of the electroweak
two-loop contributions, the fermionic contributions to
$\sin^2\theta_{\ell}^{\rm eff}$,
were calculated~\cite{Awramik:2004ge}; 
they reduce the 
intrinsic theoretical uncertainty to
$\sim 5\times 10^{-5}$.

Whether the pseudo-observable
approach will also be sufficient for Z-boson physics
at the high-luminosity GigaZ option remains to be investigated
carefully. In any case, substantial theoretical progress will be
needed to match the aimed GigaZ precision on the theoretical side,
e.g.\ for
the expected experimental accuracy in 
$\sin^2\theta_{\ell}^{\rm eff}$ of
about $1.3 \times 10^{-5}$.
A full control of observables at the two-loop level, improved by
leading higher-order effects, seems indispensable.

An important entry for the precision observables 
with a large parametric uncertainty is
the photonic vacuum polarization at the Z scale.
The hadronic part is determined via a dispersion relation from the
cross section for $\e^+ \e^- \rightarrow {\it hadrons}$ with experimental data
as input in the low-energy regime. Possible scans with the radiative-return
method require a careful theoretical treatment to reach the required
precision~\cite{Czyz:2002ft}.

\subsection{Deep-inelastic neutrino scattering}

An independent precise determination of the electroweak mixing angle
in terms of the $M_W/M_Z$ ratio has been done in deep-inelastic 
neutrino scattering off an isoscalar target
by the NuTev Collaboration~\cite{Zeller:2001hh},
yielding a deviation of about 3 standard deviations from the value
derived from the global analysis of the other precision observables. 
A new calculation of the electroweak one-loop
corrections was performed~\cite{Diener:2003ss} to investigate the stability
and size of the quantum effects,
showing that the theoretical error of the analysis in~\cite{Zeller:2001hh}
was obviously underestimated. The new theoretical result
is now being used for a re-analysis
of the experimental data (see also \cite{Arbuzov:2004zr} for another
recent recalculation of the electroweak radiative corrections).

\subsection{At the 2-loop frontier}

Although there are no complete next-to-next-to-leading (NNLO)
predictions for 
$2\to2$ scattering reactions and $1\to3$ decays (with one truly
massive leg) available yet, significant progress was reached in this
direction in recent years. 

Complete virtual two-loop amplitudes for 
(massless) Bhabha scattering \cite{Bern:2000ie} and
light-by-light scattering \cite{Bern:2001dg}
have been worked out.
Also  first steps have been made towards massive Bhabha scattering

In Ref.~\cite{Glover:2001ev} the coefficient of the 
${\cal O}(\alpha^2\log(s/m_e^2))$ 
fixed-order contribution to elastic large-angle Bhabha
scattering is derived by adapting the classification of infrared divergences
that was recently developed within dimensional regularization, and applying it
to the regularization scheme with a massive photon and electron.

The subset of factorizable corrections, resulting from one-loop
subrenormalization, is considered  
in~\cite{Fleischer:2002wa}. This requires the
evaluation of one-loop diagrams 
in arbitrary dimension $d=4-\varepsilon$.
The $\veps$-expansion
covering the orders $1/\veps$
and $\veps$,
in particular for the box graphs
needed in Bhabha scattering, was performed
in~\cite{Fleischer:2003bg}, based on the work of 
\cite{Fleischer:2003rm}~\footnote{Techniques applied 
in the latter work have also been used in 
\cite{Fleischer:2004vb}
for the analytic calculation of Higgs boson decay into two photons.}.
For the genuine two-loop QED
corrections to Bhabha scattering,
the master integrals for the box with the fermionic 
loop were calculated~\cite{bonciani1}, and
the cross section with the corrections at 
two loop resulting from these  diagrams~\cite{bonciani2}.

A complete set of master integrals for massive two-loop
Bhabha scattering has been derived in the 
meantime~\cite{czakonMI}, and for form factors with 
massless~\cite{Birthwright:2004kk}
and massive propagators~\cite{Aglietti:2004tq}.
Moreover, two-loop QCD corrections for the electroweak forward-backward
asymmetries were obtained~\cite{Bonciani:2003hc}.
Also two-loop master integrals and
form factors from virtual light fermions were
derived and applied to Higgs boson production and
decays~\cite{Aglietti:2004ki}.

\subsection{Electroweak radiative corrections at high energies}

Electroweak corrections far above the electroweak scale, e.g.\ in the
TeV range, are dominated by soft and collinear gauge-boson exchange,
leading to corrections of the form $\alpha^N\log^M(s/\MW^2)$ with
$M\le2N$. The leading terms ($M=2N$) are called Sudakov logarithms.
At the one-loop ($N=1$) and two-loop ($N=2$) level 
the leading and subleading corrections to a $2\to2$ process
at $\sqrt{s}\sim 1\TeV$ typically amount to \cite{Denner:2003wi}
\beqar
\de_{\mathrm{LL}}^{\mathrm{1-loop}} &\sim&
-\frac{\alpha}{\pi\sw^2}\log^2\bigl(\frac{s}{\MW^2}\bigr)  \simeq -26\%, 
\nn\\
\disp \de_{\mathrm{NLL}}^{\mathrm{1-loop}} &\sim&
+\frac{3\alpha}{\pi\sw^2}\log\bigl(\frac{s}{\MW^2}\bigr) \simeq 16\%,
\nn\\
\disp \de_{\mathrm{LL}}^{\mathrm{2-loop}} &\sim&
+\frac{\alpha^2}{2\pi^2\sw^4}\log^4\bigl(\frac{s}{\MW^2}\bigr) \simeq 3.5\%,
\nn\\
\disp \de_{\mathrm{NLL}}^{\mathrm{2-loop}} &\sim&
-\frac{3\alpha^2}{\pi^2\sw^4}\log^3\bigl(\frac{s}{\MW^2}\bigr) \simeq -4.2\%,
\eeqar
revealing that these corrections become significant in the high-energy phase
of a future ILC. In contrast to QED and QCD, where the Sudakov logarithms
cancel in the sum of virtual and real corrections, these terms need not
compensate in the electroweak SM for two reasons. 
The weak charges of quarks, leptons, and electroweak gauge bosons
are open, not confined, i.e.\ there is (in contrast to QCD) no need 
to average or to sum over gauge multiplets in the initial or final states
of processes. Even for final states that are inclusive with respect to the
weak charges Sudakov logarithms do not completely cancel owing to the
definite weak charges in the initial state~\cite{Ciafaloni:2000gm}.
Moreover, the large 
W- and Z-boson masses make an experimental discrimination of real W- or
Z-boson production possible, in contrast to unobservable soft-photon or
gluon emission.

In recent years several calculations of these high-energy logarithms
have been carried out in the Sudakov regime, where all kinematical
invariants $(p_i p_j)$ of different particle momenta $p_i$, $p_j$
are much larger than all particle masses.
A complete analysis of all leading and subleading logarithms at the
one-loop level can be found in~\cite{Denner:2000jv}.
Diagrammatic calculations of the leading two-loop Sudakov logarithms
have been carried out in \cite{Denner:2003wi,Beenakker:2000kb}.
Diagrammatic results on the so-called ``angular-dependent''
subleading logarithms have been presented in~\cite{Denner:2003wi}.
All these explicit results are compatible with proposed resummations
\cite{Fadin:1999bq,Kuhn:2001hz}
that are based on a symmetric SU(2)$\times$U(1) theory at high energies
matched with QED at the electroweak scale. In this ansatz,
improved matrix elements $\M$ result from lowest-order matrix elements
$\M_0$ upon dressing them with (operator-valued) exponentials,
\beq
\M \sim \M_0 \otimes \exp\left({\de_{\mathrm{ew}}}\right) \otimes
\exp\left({\de_{\mathrm{em}}}\right).
\eeq
Explicit expressions for the electroweak and electromagnetic
corrections $\de_{\mathrm{ew}}$ and $\de_{\mathrm{em}}$, which
do not commute with each other, can, for instance, be found
in \cite{Denner:2003wi}.
For $2\to2$ neutral-current processes of four massless fermions,
also subsubleading logarithmic corrections have been derived and
resummed \cite{Kuhn:2001hz}
using an infrared evolution equation that follows the pattern of QCD.
Applications to vector-boson pair production in proton--proton
collisions can be found in~\cite{Accomando:2001fn}.

In supersymmetric models the form of radiative corrections at high energies
has also been worked out for a broad class of processes 
\cite{Beccaria:2001an}. Based on 
one-loop results their exponentiation has been proposed.

\subsection{Higher-order initial-state radiation}

Photon radiation off initial-state electrons and positrons leads to
large radiative corrections of the form $\alpha^N\log^N(\Me^2/s)$.
These logarithmic corrections are universal and governed by the
DGLAP evolution equations. The solution of these equations for the
electron-photon system yields the structure functions, 
generically denoted by $\Gamma(x)$ below,
which can be used via convolution to improve hard scattering cross sections 
$\hat\sigma(p_{\Pe^+},p_{\Pe^-})$ by photon emission effects,
\beqar
\sigma(p_{\Pe^+},p_{\Pe^-}) &=& \int_0^1 \rd x_+\,\Gamma(x_+)
\int_0^1 \rd x_-\,\Gamma(x_-) \;
 \hat\sigma(x_+p_{\Pe^+},x_-p_{\Pe^-}).
\eeqar
While the soft-photon part of the structure functions 
\mbox{($x\to1$)} can be 
resummed, resulting in an exponential form, the contributions of
hard photons have to be calculated order by order in perturbation theory.
In \cite{Beenakker:1996kt} the structure functions are summarized
up to ${\cal O}(\alpha^3)$. 
\citere{Blumlein:2002bg} describes a calculation of the
(non-singlet) contributions up to ${\cal O}(\alpha^5)$ and
of the small-$x$ terms $[\alpha\log^2(x)]^N$ to all orders
(for previous calculations see papers cited in \citere{Blumlein:2002bg}).

\section{\boldmath{Radiative corrections to $2\to3,4,\dots$ processes}}
\subsection{Four-fermion final states and W-pair production}

Four-fermion final states in $e^+ e^-$ collisions, which involve electroweak
boson pair production, are of special interest since they
allow the mechanism of spontaneous symmetry 
breaking and the non-Abelian structure of the Standard Model  
 to be directly tested 
by the experiments. Moreover they provide a very important background to
most searches for new physics.

LEP2 has provided in this respect an ideal testing ground for the SM.
The W profile has been measured with great accuracy,  new bounds on 
anomalous trilinear gauge-boson couplings have been set, 
and single W, single Z, ZZ, and Z$\gamma$ cross 
sections have been determined for the first time.   
These studies will be continued with much higher statistics and
energy at a future $e^+e^-$ Linear Collider. 

In this context, the MonteCarlo four fermion generator {\tt WPHACT 2.0 } 
has been completed \cite{wph}, adapted to experimental requests and used for 
simulation of the LEP2 data  \cite{delphi}.
{\tt WPHACT 2.0} computes all Standard Model processes with four fermions 
in the final state at $e^+ e^-$ colliders, it makes use of complete,
 fully massive helicity amplitude calculations and includes the 
Imaginary Fermion Loop gauge restoring
scheme\footnote{In~\cite{Beenakker:2003va}
the incorporation of the fermion-loop corrections 
into $\Pep\Pem\to n\; {\rm fermions}$ is discussed 
in terms of an effective Lagrangian approach.}.
Thanks to these  features and new phase space mappings,
{\tt WPHACT} has been extended to all regions of phace space,
including kinematical configurations dominated by small momentum transfer and
small invariant masses
like single W, single Z, Z$\gamma ^*$ and $\gamma ^*\gamma ^*$
processes.
Special attention has been devoted to QED effects, which have a large
numerical impact, with new options for the description of Initial State 
Radiation (ISR) and of the scale dependence of the electromagnetic coupling.
Moreover, there is the possibility of including CKM mixing, and to account
for resonances in $q \bar q$ channels.

The theoretical treatment and the presently gained level in accuracy in the 
description of W-pair-mediated $4f$ production were triggered by
LEP2, as it is reviewed in \citeres{Beenakker:1996kt,Grunewald:2000ju}. 
The $\PW$ bosons are treated as resonances in the full
$4f$ processes, $\Pe^+ \Pe^- \to 4 \Pf\, (+\,\gamma)$.
Radiative corrections are split into universal and non-universal corrections.  
The former comprise leading-logarithmic corrections from
ISR, higher-order corrections included by
using appropriate effective couplings, and the Coulomb singularity.  
These corrections can be combined with the full lowest-order matrix elements
easily.
The remaining corrections are called non-universal, since they depend on
the process under investigation.
For LEP2 accuracy, it was sufficient to include these corrections
by the leading term of an expansion about the two $\PW$ poles, defining
the so-called double-pole approximation (DPA). 
Different versions of such a DPA
have been used in the literature
\cite{DPAversions,Jadach:1998tz,Denner:2000kn}.  Although several
Monte Carlo programs exist that include universal corrections, only
two event generators, {\sc YFSWW} \cite{Jadach:1998tz}
and {\sc RacoonWW} \cite{Denner:2000kn,Denner:1999gp},
include non-universal corrections, as well as the option of anomalous
gauge couplings~\cite{Denner:2001bd}.  

In the DPA approach, the W-pair cross section can be predicted within
a relative accuracy of
$\sim0.5\% (0.7\%)$ in the energy range between $180\GeV$ ($170\GeV$)
and $500\GeV$, which was sufficient for the LEP2 accuracy of
$\sim 1\%$ for energies $170{-}209\GeV$. At threshold 
($\sqrt{s}\lsim 170\GeV$), the present 
state-of-the-art prediction results from an improved Born approximation
based on leading universal corrections only, because the DPA is not
reliable there, and thus possesses an intrinsic uncertainty of
about $2\%$, which 
demonstrates the necessary theoretical improvements for the threshold region.
At energies beyond $500\GeV$, effects beyond ${\cal O}(\alpha)$,
such as the above-mentioned Sudakov logarithms at higher orders, 
become important and have to be included in predictions at
per-cent accuracy.

At LEP2, the W-boson mass has been determined by the reconstruction of 
W~bosons from their decay products with a final accuracy of about $30\MeV$.
In \cite{Jadach:2001cz} the theoretical uncertainty is estimated to be
of the order of $\sim 5\MeV$.
Theoretical improvements are, thus, desirable.

The above discussion illustrates the necessity of a full one-loop
calculation for the $\Pep\Pem\to 4f$ process
and of further improvements by leading higher-order corrections.

\subsection{Single-W production}

The single-W production process $\Pep\Pem\to\Pe\nu_\Pe\PW\to \Pe\nu_\Pe+2f$
plays a particularly important role among the $4f$ production processes
at high scattering energies. The process is predominantly initiated by
$\Pe\gamma^*$ collision,
where the photon is radiated off the electron
(or positron) by the Weizs\"acker--Williams mechanism, i.e.\ with
a very small off-shellness $q_\ga^2$. 

Consequently the cross section rises
logarithmically with the scattering energy and is of the same size
as the W-pair production cross section around $\sqrt{s}=500\GeV$;
for higher energies single-W dominates over W-pair production.

Theoretically the dominance of photon exchange at low $q_\ga^2$ poses
several complications. Technically, $q_\ga^2\to0$ means that the
electrons (or positrons) are produced in the forward direction and
that the electron mass has to be taken into account in order to
describe the cross section there. Moreover, the mere application
of $s$-dependent leading-logarithmic structure functions does not
describe the leading photon-radiation effects
properly, since ISR and final-state radiation (FSR)
show sizeable interferences for forward scattering. Thus, 
the improvement of lowest-order calculations by leading radiation 
effects is more complicated than for $s$-channel-like processes.
Finally, the running of the electromagnetic coupling 
$\alpha(q_\ga^2)$ has to be evaluated in the region of small
momentum transfer ($q_\ga^2<0$) where the fit for the hadronic part 
of the photon vacuum polarisation~\cite{Jegerlehner:2001ca}
should be used. 

The Monte Carlo generator {\sc KoralW} \cite{Jadach:1998gi}
has been updated to include the ISR-FSR interference effects
as well as the proper running of $\alpha(q_\ga^2)$.
Therefore, this program now has reached a level of accuracy
similar to the other state-of-the-art programs for single-W production:
{\sc Grc4f} \cite{Fujimoto:1996qg},
{\sc Nextcalibur} \cite{Berends:2000gj},
{\sc Swap} \cite{Montagna:2000ev},
{\sc Wphact} \cite{Accomando:1996es,wph}, and
{\sc Wto} \cite{Passarino:1996rc}.
It should be kept in mind that none
of these calculations includes non-universal electroweak
corrections, leading to a theoretical uncertainty of about
$\sim5\%$ in cross-section predictions. Although the final
solution for a high-energy Linear Collider
certainly requires a full
${\cal O}(\alpha)$ calculation of the $4f$-production process,
a first step of improvement could be done by a careful
expansion about the propagator poles of the photon and W~boson.
The electroweak ${\cal O}(\alpha)$ corrections to the process
$\Pe\ga\to\nu_\Pe\PW$, which are known \cite{Denner:1992vg},
represent a basic building block in this calculation.
\looseness -1

\subsection{Progress for 
multi-particle production processes}

One-loop integrals become more and more cumbersome if the number $N$ of
external legs in diagrams increases. For $N>4$,
however, not all external momenta are linearly independent because of the
four-dimensionality of space-time.
As known for a long time \cite{Me65}, this fact opens the possibility 
to relate integrals with $N>4$ to integrals with $N\le4$.
In recent years, various techniques for actual evaluations of
one-loop integrals with $N=5,6$ have been worked out
\cite{Fleischer:1999hq,Denner:2002ii} 
(see also references therein for older methods and results).
The major complication in the treatment of $2\to3$ processes at one loop
concerns the numerical evaluation of tensor 5-point integrals; in particular,
the occurrence of inverse Gram determinants in the usual
Passarino--Veltman reduction to scalar integrals leads
to numerical instabilities at the phase-space boundary. A possible
solution to this problem was worked out in \cite{Denner:2002ii}
where the known direct reduction 
\cite{Me65} of scalar 5-point to 4-point integrals
was generalized to tensor integrals, thereby avoiding the occurrence
of leading Gram determinants completely.
More work on one-loop $N$-point integrals can be found 
in~\cite{Giele:2004iy}.

In the evaluation of real corrections, such as bremsstrahlung,
a proper and numerically stable separation of infrared (soft and collinear)
divergences represents one of the main problems.
In the phase-space slicing approach 
(see \cite{Harris:2001sx} and references therein)
the singular regions are excluded from the ``regular'' 
phase-space integration by small cuts on energies, angles, or
invariant masses. Using factorization properties, the integration over
the singular regions can be done in the limit of infinitesimally
small cut parameters. 
The necessary fine-tuning of cut parameters is avoided in so-called
subtraction methods 
(see \cite{Catani:1997vz,Dittmaier:2000mb,Phaf:2001gc} 
and references therein),
where a specially tuned auxiliary function is subtracted from the 
singular integrand in such a way that the resulting integral is regular.
The auxiliary function has to be chosen simple enough, so that the
singular regions can be integrated over analytically. 
In \cite{Catani:1997vz} the so-called ``dipole subtraction approach''
has been worked out for massless QCD. 
and subsequently extended for  photon emission off massive
fermions~\cite{Dittmaier:2000mb} and for QCD with massive
quarks~\cite{Phaf:2001gc}.

Applications were performed for 
complete one-loop calculations of electroweak radiative corrections
for specific $2\to3$ processes of special interest for
a future ILC,
$\Pep\Pem\to\nu\bar\nu\PH$ \cite{Belanger:2002ik,Denner:2003yg} and
$\Pep\Pem\to\Pt\bar\Pt\PH$ \cite{You:2003zq,Belanger:2003nm,eetth}.
In \cite{Denner:2003yg,You:2003zq,eetth} the 
technique \cite{Denner:2002ii} for treating tensor 5-point integrals
was employed. 
While \cite{Belanger:2002ik,You:2003zq,Belanger:2003nm} make
use of the slicing approach for treating soft-photon emission,
the results of \cite{Denner:2003yg,eetth} have been obtained by
dipole subtraction and checked by phase-space slicing
for soft and collinear bremsstrahlung.
Analytic results for the one-loop corrections to 
$\Pep\Pem\to\nu\bar\nu\PH$ are provided
in~\cite{Jegerlehner:2002es}.

The Yukawa coupling of the top quark could be measured at a future
ILC with high energy and luminosity at the level of $\sim5\%$ 
\cite{Aguilar-Saavedra:2001rg} by analyzing the process $\eetth$.
A thorough prediction for this process, thus, has to control QCD
and electroweak corrections. 
Results on the electroweak
$\Oa$ corrections of \citeres{Belanger:2003nm,eetth} show
agreement within $\sim 0.1\%$.
The results of the previous calculation \cite{You:2003zq}
roughly agree with the ones of \citeres{Belanger:2003nm,eetth} at
intermediate values of $\sqrt{s}$ and $\MH$, but are at variance
at high energies (TeV range) and close to threshold (large $\MH$).

\section{Event generators for multi-particle final states}

\subsection{Multi-purpose generators at parton level}

The large variety of different final states for multi-particle
production renders multi-purpose Monte Carlo event generators rather 
important, i.e.\ generators that deliver
an event generator for a user-specified (as much as possible)
general final state based on full lowest-order amplitudes.
As results, these tools yield lowest-order predictions for
observables, or more generally Monte Carlo samples of events,
that are improved by universal radiative corrections, such as
initial-state radiation at the leading-logarithmic level or 
beamstrahlung effects.
Most of the multi-purpose generators are also interfaced to parton-shower
and hadronization programs.
The generality renders these programs, however, rather complex devices
and, at present, they are far from representing tools for high-precision
physics, because non-universal radiative corrections are not taken 
into account in predictions.

The following multi-purpose generators for multi-parton production, 
including program packages for the matrix-element evaluation, are available:
\begin{itemize}
\item
{\sc Amegic} \cite{Krauss:2001iv}: 
Helicity amplitudes are automatically generated by the program for
the SM, the MSSM, and some new-physics models. Various interfaces
(ISR, PDFs, beam spectra, {\sc Isajet}, etc.) are supported.
The phase-space generation was successfully tested for up to
six particles in the final state.
\item
{\sc Grace} \cite{Ishikawa:1993qr}: The amplitudes are delivered
by a built-in package, which can also handle SUSY processes. The
phase-space integration is done by {\sc BASES} \cite{Kawabata:1995th}.
Tree-level calculations have been performed for up to 
(selected) six-fermion final states. The extension of the system
to include one-loop corrections is the {\sc Grace-Loop} \cite{Fujimoto:zx}
program.
\item
{\sc Madevent} \cite{Maltoni:2002qb} + {\sc Madgraph} \cite{Stelzer:1994ta}:
The {\sc Madgraph} algorithm can generate tree-level matrix elements
for any SM process (fully supporting particle masses), but a practical 
limitation is 9,999 diagrams.
In addition, {\sc Madgraph} creates {\sc Madevent}, an event generator for 
the requested process. 
\item
{\sc Phegas} \cite{Papadopoulos:2000tt} + {\sc Helac} \cite{Kanaki:2000ey}:
The {\sc Helac} program delivers amplitudes for all SM processes 
(including all masses). The phase-space integration done by {\sc Phegas}
has been tested for selected final states with up to seven particles.
Recent applications concern channels with six-fermion final 
states~\cite{Gleisberg:2003bi}.
\item
{\sc Whizard} \cite{Kilian:2001qz} 
+ {\sc Comphep} \cite{Pukhov:1999gg} / {\sc Madgraph} \cite{Stelzer:1994ta}
/ {\sc O'mega} \cite{Moretti:2001zz}:
Matrix elements are generated by an automatic interface to (older versions of)
{\sc Comphep}, {\sc Madgraph}, and (the up-to-date version of) {\sc O'mega}.
Phase-space generation has been tested for most $2\to6$ and some
$2\to8$ processes; unweighted events are supported, and a large variety
of interfaces (ISR, beamstrahlung, {\sc Pythia}, PDFs, etc.) exists.
The inclusion of MSSM amplitudes ({\sc O'mega}) and improved
phase-space generation ($2\to6$) are work in progress.
\item
{\sc Alpgen}~\cite{alpgen} is a specific code for computing
the perturbative part of 
observables in high energy hadron--hadron collisions, which 
require a convolution of the perturbative quantities 
with structure or fragmentation functions that account for 
non-perturbative effects.
\end{itemize}
Tuned comparisons of different generators, both at parton and 
detector level, are extremely important, but become more and more
laborious owing to the large variety of multi-particle final states.
Some progress to a facilitation and automization of comparisons 
are made by MC-tester project \cite{Golonka:2002rz} and 
Java interfaces \cite{Ronan:2003wq}.

\subsection{Event generators and results for $\Pep\Pem\to6f$}

Particular progress was reached in recent years in the description
of six-fermion production processes. Apart from the multi-purpose generators
listed in the previous section, also dedicated Monte Carlo programs
and generators have been developed for this class of processes:
\begin{itemize}
\item
{\sc Sixfap} \cite{Montagna:1997dc}: Matrix elements are provided
for all $6f$ final states (with finite fermion masses), including
all electroweak diagrams. The generalization to QCD diagrams and the
extension of the phase-space integration for all final states is in
progress.
\item
{\sc eett6f} \cite{Kolodziej:2001xe}: Only processes relevant for
$\Pt\bar\Pt$ production are supported (a new version includes
$\Pe^\pm$ in the
final state and QCD diagrams); finite fermion masses are possible. 
\item
{\sc Lusifer} \cite{Dittmaier:2002ap}: All $6f$ final states are
possible, including QCD diagrams with up to four quarks; 
representative results for all these final states have been presented.
External fermions are massless.
An unweighting algorithm and an interface to {\sc Pythia} are available.
\item
{\sc Phase} \cite{Accomando:2004my}: All Standard Model process with six fermions
in the final state at the LHC. It employs the full set of tree-level
Feynman diagrams, taking into account a finite mass for the b quark. An interface 
to hadronization is provided.
\end{itemize}
A comparison of results
for some processes $\Pep\Pem\to 6f$ relevant for $\Pt\bar\Pt$ production
for massless external fermions
reveals good agreement between the various programs~\cite{Dittmaier:2003sc},
where minor differences are presumably due to the different treatments of the 
bottom-quark Yukawa coupling, which is neglected in some cases.

A tuned comparison of results obtained with {\sc Lusifer} and
{\sc Whizard} for a large survey of $6f$ final states has been
presented in \citere{Dittmaier:2002ap}.

\section{Other developments}

\subsection{Automatization of loop calculations}

Once the necessary techniques and theoretical subtleties of a 
perturbative calculation are settled, to carry out the actual
calculation is an algorithmic matter. Thus, an automization of
such calculations is highly desirable, in order to facilitate
related calculations. Various program packages
have been presented in the literature for automatized tree-level,
one-loop, and multi-loop calculations. A comprehensive
overview can, e.g., be found in~\cite{Harlander:1998dq};
in the following we have to restrict ourselves to a selection of
topics, where emphasis is put on electroweak aspects.

The generation of Feynman
graphs and amplitudes is a combinatorial problem
that can be attacked with computer algebra.
The program packages {\sc FeynArts}~\cite{Kublbeck:1990xc}
(which has been extended in~\cite{Hahn:2001rv} for the MSSM)
and {\sc Diana}~\cite{Tentyukov:1999is}
(based on {\sc Qgraf}~\cite{Nogueira:1991ex}) 
are specifically designed for this task; also the 
{\sc Grace-Loop} \cite{Fujimoto:zx} system automatically generates
Feynman diagrams and loop amplitudes. 
Moreover, the task of calculating
virtual one-loop and the corresponding real-emission corrections
to $2\to 2$ scattering reactions is by now well understood.
Such calculations are widely automated in the packages
{\sc FormCalc} combined with {\sc LoopTools}~\cite{Hahn:1998yk},
and {\sc Grace-Loop}~\cite{Fujimoto:zx}.

An illustrating example was provided for
the differential cross section for $\Pep\Pem\to\Pt\bar\Pt$
in lowest order as well as including electroweak ${\cal O}(\alpha)$
corrections.
A program {\sc FA}+{\sc FC}~\cite{Fleischer:2002rn} 
was obtained from the output of the
{\sc FeynArts} and {\sc FormCalc} packages and makes use of the
{\sc LoopTools} library for the numerical evaluation.
Another program,
{\sc Topfit}~\cite{Fleischer:2002rn,Fleischer:2003kk},
was developed from an algebraic reduction of Feynman graphs 
(delivered from {\sc Diana}) within {\sc Form}; for the
numerics {\sc LoopTools} is partially employed. 
A completely independent approach has been made by the
{\sc Sanc} project~\cite{Andonov:2002jg}. 
The {\sc Sanc} program contains another 
independent calculation of the ${\cal O}(\alpha)$
corrections to $\Pep\Pem\to\Pt\bar\Pt$, the results of which are also
included in~\cite{Fleischer:2002rn}.
More details on comparisons,
including also other fermion flavours, 
can be found in \cite{Dittmaier:2003sc,Fleischer:2002rn,Hahn:2003ab}.
The agreement between the numerical results at the level of 10 digits 
reflects the enormous
progress achieved in recent years in the automatization of
one-loop calculations.

\def\aitalc{{\sc \textit{a}{\r{\i}}\raisebox{-0.14em}{T}alc}}
The one-loop calculation for the process 
$\Pep\Pem\to\Pt\bar\Pt (\gamma)$ including
hard bremsstrahlung was originally performed in~\cite{Fleischer:2003kk} 
without full automatization; it was repeated in later course
(apart from the hard bremsstrahlung part)
as an example for automatization in~\cite{Fleischer:2004ah} .
The extension of {\sc Diana} towards 
full automatization in terms of the package {\aitalc} 
is a new development~\cite{Fleischer:2004ah,Lorca:2004dk};
automatization of the full calculation is performed
including renormalization and the creation and running of a FORTRAN code. 
Applications
to the calculation of the processes 
$\Pep\Pem\to f \bar{f} (\gamma)$ for 
various final
fermions: $t,b,\mu,e$ and also $s\bar{b},c\bar{t},\mu\bar{\tau}$ are 
performed.
For further work in automatization see \cite{Tentyukov:2003bx,ACAT03}.

\subsection{Numerical approaches to loop calculations}

Most of the various techniques of performing loop calculations share
the common feature that the integration over the loop momenta is 
performed analytically. This procedure leads to complications
at one loop if five or more external legs are involved, since
both speed and stability of programs become more and more 
jeopardized. At the two-loop level, already the evaluation of 
self-energy and vertex corrections can lead to extremely complicated
higher transcendental functions that are hard to evaluate numerically.

An idea to avoid these complications is provided by a more or less
purely numerical evaluation of loop corrections. There are two main
difficulties in this approach. Firstly, the appearing ultraviolet and
infrared divergences have to be treated and canceled carefully. 
Secondly, even finite loop integrals require a singularity handling
of the integrand near particle poles, where Feynman's
$\ri\eps$ prescription is used as regularization.

In \cite{Passarino:2001wv}-\cite{Ferroglia:2003yj}
a method for a purely numerical evaluation of
loop integrals is proposed. Each integral is parametrized with 
Feynman parameters and subsequently rewritten with partial
integrations. The final expression consists of a quite simple part 
containing the singular terms and another more complicated looking
part that can be integrated numerically. 
The actual application of the method to a physical process is still
work in progress.

There are five papers in a series devoted to the numerical evaluation of
multi-loop, multi-leg Feynman diagrams. In~\cite{Passarino:2001wv}
the general strategy is outlined and
in~\cite{Passarino:2001jd} a complete list of results is given
for two-loop functions with two external legs, including their
infrared divergent on-shell derivatives. Results for one-loop multi-leg
diagrams are shown in~\cite{Ferroglia:2002mz} and additional material can
be found in~\cite{Ferroglia:2002yr}. Two-loop three-point functions for
infrared convergent configurations are considered in~\cite{Ferroglia:2003yj},
where numerical results can be found.

\citere{Ferroglia:2002mz} 
presents a detailed investigation of the algorithms, based
on the Bernstein-Tkachov (BT) theorem~\cite{Tkachov:1997wh},
which form the basis
for a fast and reliable numerical integration of one-loop multi-leg (up to
six in this paper) diagrams. The rationale for this work is represented by 
the need of encompassing a number of problems that one encounters in 
assembling a calculation of some complicated process, e.g.\ full one-loop 
corrections to $\Pep\Pem \to 4\,$fermions. Furthermore, in any attempt to 
compute physical observables at the two-loop level, we will have to include 
the one-loop part, and it is rather obvious that the two pieces should be 
treated on equal footing. 

All algorithms that aim to compute Feynman diagrams numerically are based on
some manipulation of the original integrands that brings the final answer
into something smooth. This has the consequence of bringing the original
(Landau) singularity of the diagram into some overall denominator and,
usually, the method overestimates the singular behavior around some
threshold. In these regions an alternative derivation is needed. Instead of
using the method of asymptotic expansions, a novel algorithm is introduced 
based on a Mellin-Barnes decomposition of the singular integrand,
followed by a sector decomposition that allows one to write the Laurent
expansion around threshold. 

Particular care has been devoted to analyze those situations where a 
sub-leading singularity may occur, and to properly account for those cases 
where the algorithm cannot be applied because the corresponding BT factor 
is zero although the singular point in parametric space does not belong to 
the integration domain.

Finally, a description of infrared divergent one-loop virtual
configurations is given
in the framework of dimensional regularization: here both the 
residue of the infrared pole and the infrared finite remainder are cast into 
a form that can be safely computed numerically.
The collection of formulas that cover all corners of phase space have been
translated into a set of FORM codes and the output 
has been used to create a FORTRAN code whose technical description will be 
given elsewhere. 

\citere{Actis:2004bp} addresses the problem of deriving a judicious and 
efficient way to deal with tensor Feynman integrals, namely those integrals 
that occur in any field theory with spin and non trivial structures for the 
numerators of Feynman propagators. This paper forms a basis for a
realistic calculation of physical observables at the two-loop level.

The complexity of handling two-loop tensor integrals is reflected in the
following simple consideration: the complete treatment of one-loop tensor
integrals was confined to the appendices of~\cite{Passarino:1979jh}, while
the reduction of general two-loop self-energies to standard scalar integrals
already required a considerable fraction of~\cite{Weiglein:hd}; the
inclusion of two-loop vertices requires the whole content of this paper.
The past experience in the field has shown the encompassed convenience of
gathering in one single place the complete collection of results needed for
a broad spectrum of applications.
In recent years, the most popular and quite successful tool in dealing with
multi-loop Feynman diagrams in QED/QCD (or in selected problems in different
models, characterized by a very small number of scales), has been the
Integration-By-Parts Identities method~\cite{'tHooft:1972fi}. 
However, reduction to a set of master
integrals is poorly known in the enlarged scenario of multi-scale
electroweak physics.

In~\cite{delAguila:2004nf} another new method is presented in which almost
all the work can be performed numerically: 
one-loop tensor integrals  are
numerically reduced to the standard set of one-loop scalar functions and 
any amplitude is calculated simply contracting
the numerically computed tensor integrals with the external 
tensors.
To this aim, a recursion relation is introduced 
that links high-rank tensor integrals to lower-rank ones.
Singular kinematical configurations give a smoother behaviour than in other 
approaches because, 
at each level of iteration, 
only inverse square roots of Gram determinants appear.

\section{Electroweak effects in hadronic processes}

In Refs.~\cite{Maina:2004ve,Maina:2004rb,Maina:2004dm,Maina:2003is,Kuhn:2004em} 
(see also \cite{Dobbs:2004bu}) it was 
proved the importance of electroweak one-loop 
corrections to hadronic observables, such as $b\bar b$,
`prompt photon + jet' and `$Z$ + jet' production at 
Tevatron and LHC and jet and $b\bar b$ production in linear colliders,
which can compete in size with QCD corrections.
Their inclusion in
experimental analyses is thus important, especially in view of 
searches for new physics. In case of `$Z$ + jet' production
they can rise to ${\cal O}(15-20\%)$ at large transverse momentum at
the LHC, while being typically half the size in case of `photon + jet'
production. As these two channels are the contributors to the Drell-Yan
process, and since the latter is envisaged to be used as one of the means
to measure the LHC luminosity, it is clear that neglecting them in
experimental analyses would spoil the luminosity measurements.

Ref.~\cite{Maina:2002wz} emphasised the importance of 
NLO electroweak effects
in three-jet production in $\e^+\e^-$ scattering at the $Z$-pole (SLC, LEP
and GigaZ), showing typical corrections of  ${\cal O}(2-4\%)$ (e.g.
in jet-rates and thrust), comparable to the SLC and LEP experimental accuracy
and certainly larger than the one expected at GigaZ. 
They also introduce sizable parity-violating effects into the fully 
differential structure of three-jet events in presence of beam polarisation,
which are of relevance as background to 
new physics searches in SLC and GigaZ data.

The complete set of electroweak $O(\alpha)$ corrections to the Drell--Yan-like
production of $Z$ and $W$ bosons have been 
studied in~\cite{Brein,Dittmaier:2001ay,Baur:2004ig}. 
These corrections are phenomenologically relevant both
at the Tevatron and the
LHC. It is shown that the pole expansion yields a good 
description of resonance
observables, but it is not sufficient for the high-energy tail of
transverse-momentum distributions, relevant for new-physics searches.
{\sc Horace} and {\sc Winhac} are
Monte Carlo event generators~\cite{CarloniCalame:2004qw}, developed 
for single W production and decay, which in their current versions
include higher-order QED corrections in leptonic W decays,
a crucial entry for precision determination of the W mass and width
at hadron colliders.

Production of vector-boson pairs is an important probe for potential
non-standard anomalous gauge couplings.
In order to identify possible deviations from the SM predictions,
an accurate knowledge of the electroweak higher-order contributions
is mandatory as well, in particular for large transverse momenta.
A complete electroweak one-loop calculation was performed
for $\gamma Z$ production~\cite{Meier}; for other processes
like $\gamma W, ....$ the large logarithms in the Sudakov regime
were derived~\cite{Accomando:2001fn}.

A further aspect that should be recalled is that weak corrections naturally
introduce parity-violating effects in observables, detectable through
asymmetries in the cross-section.
These effects are further enhanced if polarisation
of the incoming beams is exploited, such as at RHIC-Spin
\cite{Bourrely:1990pz,Ellis:2001ba} 
and will be used to measure polarised structure functions.

\section{Conclusions}

During the recent years
there has been 
continuous progress in the development of new techniques and in
making precise predictions for electroweak physics at future colliders.
However, to be prepared for the LHC and
a future $\Pep\Pem$ linear collider
with high energy and luminosity, an enormous
amount of work is still ahead of us. 
Not only technical challenges, also
field-theoretical issues such as
renormalization, the treatment of unstable particles, etc.,
demand a higher level of understanding. 
Both loop calculations as well as the descriptions
of multi-particle production processes with the help of Monte Carlo
techniques require and will profit from further improving
computing devices.
It is certainly out of question that the list of challenges 
and interesting issues could be continued at will. 
Electroweak physics at
future colliders will be a highly exciting issue.


\end{document}